\documentclass[english,manuscript,superscriptaddress]{revtex4}
\usepackage[T1]{fontenc}
\usepackage[latin9]{inputenc}
\usepackage{graphicx}
\usepackage{amssymb}

\providecommand{\tabularnewline}{\\}

\usepackage{babel}

\begin{document}

\title{A Search for leptophilic $Z_{l}$ boson at future linear colliders}

\author{S.O. Kara}

\email{sokara@science.ankara.edu.tr}

\affiliation{Ankara University, Physics Department, Ankara, Turkey}

\author{M. Sahin}

\email{m.sahin@etu.edu.tr}

\affiliation{TOBB University of Economics and Technology, Physics Division, Ankara,
Turkey}

\author{S. Sultansoy}

\email{ssultansoy@etu.edu.tr}

\affiliation{TOBB University of Economics and Technology, Physics Division, Ankara,
Turkey}

\affiliation{Institute of Physics, National Academy of Sciences, Baku, Azerbaijan}

\author{S. Turkoz}

\email{turkoz@science.ankara.edu.tr}

\affiliation{Ankara University, Physics Department, Ankara, Turkey}
\begin{abstract}
We study the possible dynamics associated with leptonic charge in
future linear colliders. Leptophilic massive vector boson, $Z_{l}$,
have been investigated through the process $e^{+}e^{-}$$\rightarrow$$\mu^{+}\mu^{-}$.
We have shown that ILC and CLIC will give opportunity to observe Z$_{l}$
with masses up to the center of mass energy if the corresponding coupling
constant $g_{l}$ exceeds $10^{-3}$. 
\end{abstract}
\maketitle

\section*{1. Introduction}

Historically baryon and lepton number's conservations had been proposed
to explain non observation of certain processes such as $p\rightarrow e^{+}\gamma$
etc. Even though these conserved quantities are not the outcome of
the standard model (SM), they can be incorporated into the SM as accidental
global symmetries. It is natural to consider possible gauging of these
numbers in analogy with gauging of electric charge in QED. The gauging
of the baryon and lepton numbers has a long history. In 1955 Lee and
Yang proposed massless baryonic {}``photon'' \cite{Lee}, later
in 1969 Okun considered massless leptonic {}``photon'' \cite{Okun1}
in analogy with the baryonic photon. However, the experiments on the
equivalence of inertial and gravitational masses have put very strong
limits on the strength of corresponding coupling constants, namely,
$\alpha_{B}<10^{-47}$ and $\alpha_{L}<10^{-49}$ for baryonic and
leptonic photons respectively. Comparing these with $\alpha_{EM}\approx10^{-2}$
has led to the conclusion that massless baryonic and leptonic photons
are not exist in nature. The interest on leptonic photon has been
revived in 1995 \cite{Ciftci} with the consideration of possible
compensation of leptonic charge of matter by relic anti-sneutrinos
foreseen by the standard big bang theory and SUSY. It was shown that
the available experimental data admit the additional range for the
leptonic interaction constant namely $10^{-38}<\alpha_{L}<10^{-14}$
as a consequence of this compensation. This result has led to a number
of studies on the subject \cite{Okun2,Okun3,Martemyanov,Gninenko,Ilyin,Akkus,Dolgov,Brockway}.

It should be noted that there is no compensation mechanism for baryonic
charge. For this reason massive baryonic vector boson has been proposed
for baryonic charge \cite{Carone1,Carone2} and it was shown {}``that
a new gauge boson $\gamma_{B}$ coupling only to baryon number is
phenomenologically allowed, even if $m_{B}<m_{Z}$ and $\alpha_{B}\thickapprox0.2$''.
On the other hand gauging of B-L \cite{Buchmuller,Shaaban} is natural
in the framework of Grand Unification Theories. Manifestations of
the $Z^{'}$ boson of the minimal B-L model at future linear colliders
and LHC have been considered in recent paper \cite{Basso}. Table
1 reflects today's status of B, L and B-L studies. As it seen from
the Table the massive leptonic boson, as well as massless B-L boson
have not been considered so far.

\begin{table}
\begin{tabular}{|c|c|c|}
\hline 
 & massless & massive\tabularnewline
\hline
\hline 
B & + & +\tabularnewline
\hline 
L & + & -\tabularnewline
\hline 
B-L & - & +\tabularnewline
\hline
\end{tabular}

\caption{Status of the studies related to B, L and B-L gauge bosons: plus sign
indicates the subject have been considered. }

\end{table}

In this paper we have considered phenomenology of massive $U(1)$
boson coupled to lepton charge (leptophilic/quarkophobic $Z_{l}$).
In section 2, the model has been formulated. Production of leptophilic
$Z_{l}$ at future lepton colliders (ILC/CLIC) is analyzed in section
3. In the final section the results obtained are summarized.

\section*{2. The Model}

To gauge the leptonic quantum number in our model we add a new $U_{l}^{'}(1)$
gauge symmetry to standard model (SM) gauge group $(SU_{C}(3)\times SU_{W}(2)\times U_{Y}(1))$.
It should be noted that the experimental discovery of neutrino oscillations
\cite{PDG2010} has invalidated the idea of conservation of electron,
muon and tau lepton charges individually. In our model we consider
single lepton charge which is the same for $e$, $\mu$, $\tau$ and
corresponding neutrinos. In the model, the interaction of the electroweak
vector bosons with fermions and Higgs fields is introduced through
the following replacement in the free fields Lagrangian:

\begin{equation}
\partial{}_{\mu}\rightarrow D_{\mu}=\partial{}_{\mu}-ig_{2}\mathbf{T\cdot A_{\mu}}-ig_{1}\frac{Y}{2}B_{\mu}-ig_{l}a_{l}B_{\mu}^{'}\end{equation}
where $g_{2}$, $g_{1}$ and $g_{l}$ are interaction constants, \textbf{$\mathbf{T}$}
is an isospin operator of a corresponding multiplet of fermionic or
Higgs fields, $Y$ is hypercharge and $a_{l}$ is lepton charge of
the corresponding multiplet, \textbf{$\mathbf{A}_{\mu}$}, $B_{\mu}$,
$B_{\mu}^{'}$ are gauge fields. Higgs field with lepton charge must
be added to provide mass to leptophilic $Z_{l}$ boson which in our
model coincides with $B_{\mu}^{'}$ vector field. Interaction Lagrangian,
obeying the $SU_{C}(3)\times SU_{W}(2)\times U_{Y}(1)\times U_{l}^{'}(1)$
gauge symmetry, can be decomposed as:

\begin{equation}
L=L_{SM}+L^{'}\end{equation}
 where $L_{SM}$ is standard model Lagrangian and $L^{'}$ is given
by:

\begin{equation}
L^{'}=\frac{1}{4}F_{\mu\nu}^{'}F^{\mu\nu'}+g_{l}J_{lep}^{\mu}B_{\mu}^{'}+(D_{\mu}\Phi)^{\dagger}(D^{\mu}\Phi)+\mu^{2}\left|\Phi\right|^{2}-\lambda\left|\Phi\right|\end{equation}
 where

\begin{equation}
F_{\mu\nu}^{'}=\partial_{\mu}B_{\nu}^{'}-\partial_{\nu}B_{\mu}^{'}\end{equation}
 is field strength tensor,

\begin{equation}
J_{lep}^{\mu}=\sum_{l}a_{l}[\bar{\nu_{l}}\gamma^{\mu}\nu_{l}+\bar{l}\gamma^{\mu}l]\end{equation}
 is leptonic current interacting with leptophilic $Z_{l}$ , $\Phi$
is singlet complex scalar Higgs field. To avoid the triangular anomalies
the following condition should be satisfied in our model

\begin{equation}
\sum_{l}a_{l}=0.\end{equation}

As mentioned before, the experimental data on neutrino oscillations
requires the same leptonic charge for $e$, $\mu$, $\tau$ and corresponding
neutrinos $(a_{e}=a_{\mu}=a_{\tau}=1)$. Therefore, additional fermion
families are needed to satisfy the condition (6). It is known that
recent precision electroweak data allows the existence of the fourth
SM family \cite{Maltoni,He,Vysotsky,Novikov1,Bulanov,Novikov2,Alwall,Kribs,Bobrowski,Chanowitz,Hashimoto,Ebberhardt,Cobanoglu,opucem,Erler,Sahin}.
In this case to satisfy the above condition in our model we take lepton
charge of the fourth family leptons is equal to $-3$ \cite{Carone2,Perez}.

\section*{3. Production of the leptophilic $Z_{l}$ boson at future linear
colliders }

For numerical calculations we implement the Lagrangian (3) into the
CALCHEP Simulation Program \cite{Pukhov}. In new generation linear
colliders initial state radiation (ISR) and beamstrahlung (BS) will
be important. Therefore, we use beam design parameters given in table
2 \cite{Brau,Braun,CLIC}.

\begin{table}
\begin{tabular}{|c|c|c|c|}
\hline 
Collider Parameters  & ILC  & \multicolumn{2}{c||}{CLIC}\tabularnewline
\hline
\hline 
E($\sqrt{S}$) TeV  & $0.5$  & $0.5$  & $3$ \tabularnewline
\hline 
L($10^{34}$ $cm^{-2}$$s^{-1}$)  & $2$ & $2.3$ & $5.9$\tabularnewline
\hline 
$N$($10^{10}$)  & $2$ & $0.68$ & $0.372$\tabularnewline
\hline 
$\sigma_{x}$(nm)  & $640$ & $202$ & $45$\tabularnewline
\hline 
$\sigma_{y}$(nm)  & $5.7$ & $2.3$ & $1$\tabularnewline
\hline 
$\sigma_{z}$($\mu m$)  & $300$ & $44$ & $44$\tabularnewline
\hline
\end{tabular}

\caption{Main parameters of ILC and CLIC. Here N is the number of particles
in bunch, $\sigma_{x}$ and $\sigma_{y}$ are RMS transverse beam
sizes at Interaction Points (IP), $\sigma_{z}$ is the RMS bunch length.}

\end{table}

Before proceeding to calculations we ought to define the parameter
space of our model compliant with existing experimental constraints.
Limits from precision electroweak data on different kinds of $Z^{'}$
bosons have been obtained in \cite{Cacciapaglia,Aguila}. We decided
to use here the conservative constraint from \cite{Cacciapaglia}:

\begin{equation}
\frac{M_{Z_{l}}}{g_{l}}\geq7\, TeV.\end{equation}
 For given mass values the upper bounds of coupling constants obeying
constraint (7), are displayed in table 3.

\begin{table}
\begin{tabular}{|c|c|}
\hline 
$M_{Z_{l}}$(TeV)  & $g_{l}$\tabularnewline
\hline
\hline 
$0.5$ & $0.07$\tabularnewline
\hline 
$1.0$ & $0.14$\tabularnewline
\hline 
$1.5$ & $0.21$\tabularnewline
\hline 
$2.0$ & $0.28$\tabularnewline
\hline 
$2.5$ & $0.35$\tabularnewline
\hline 
$3.0$ & $0.42$\tabularnewline
\hline
\end{tabular}

\caption{Upper bounds of the coupling constant for different values of $Z_{l}$
mass. }

\end{table}

In all calculations we have done, our signal process is $e^{+}e^{-}\rightarrow\gamma,\, Z,\, Z_{l}\rightarrow\mu^{+}\mu^{-}$
and background process is $e^{+}e^{-}\rightarrow\gamma,\, Z\rightarrow\mu^{+}\mu^{-}$.
This process is chosen because it is more clean than other possible
processes: Final state containing $e^{+}e^{-}$ pair has a huge background
(i.e. due to bhabha scattering); $\tau^{+}\tau^{-}$ pair will complicate
the signal due to $\tau$ decays ; $\overline{\nu}\nu$ pair final
states are unobservable.

In Figure 1(2) cross section versus $M_{Z_{l}}$ at ILC (CLIC with
$\sqrt{S}=0,5$ TeV) is plotted for different values of coupling constant.
It is seen that the signal is well above the SM background even for
small values of $g_{l}$. For the mass values less than $0.5$ TeV
the signal is above the background due to positive interferences between
$\gamma$, $Z$ and $Z_{l}$. Comparing Figure 1 and 2 show that ILC
is advantageous for $M_{Z_{l}}\approx0.5$ TeV, whereas for smaller
values of $M_{Z_{l}}$ CLIC gives larger difference between signal
and background.

\begin{figure}
\includegraphics[scale=0.7]{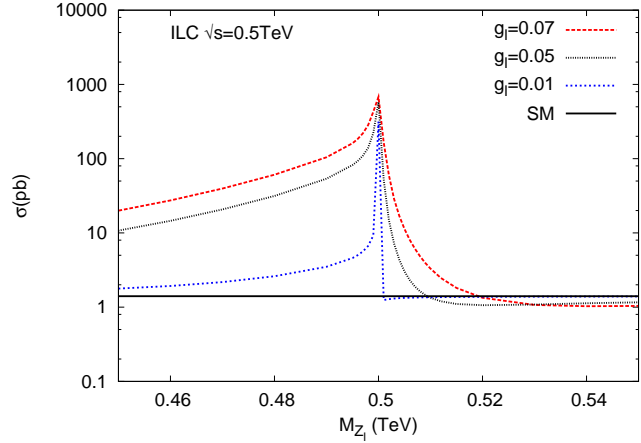}

\caption{Cross section versus $Z_{l}$ mass for different coupling values and
SM background at ILC with $\sqrt{S}=0.5$ TeV}

\end{figure}

\begin{figure}
\includegraphics{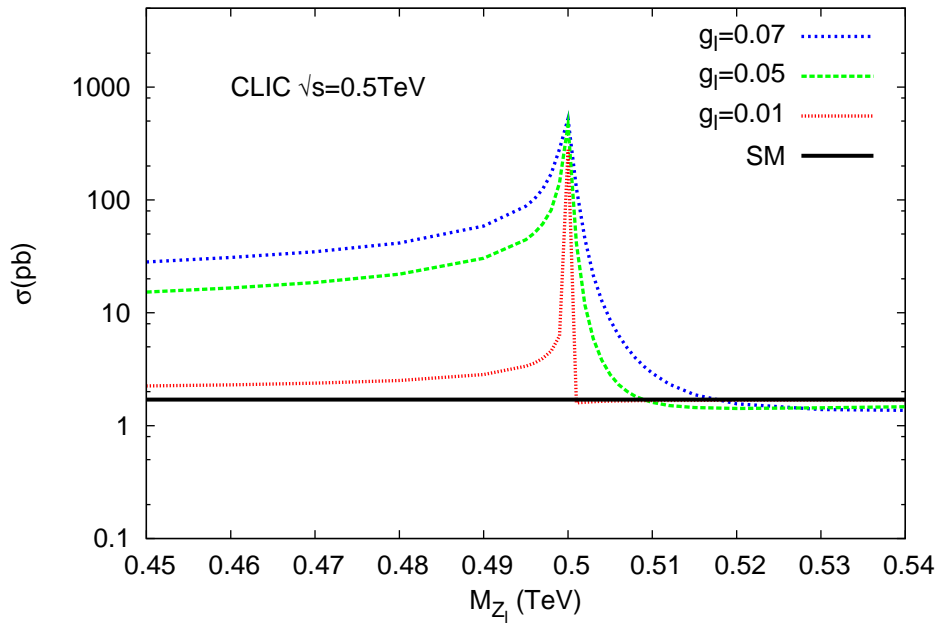}

\caption{Cross section versus $Z_{l}$ mass for different coupling values and
SM background at CLIC with $\sqrt{S}=0.5$ TeV.}

\end{figure}

Cross section versus $M_{Z_{l}}$ for CLIC with $\sqrt{S}=3$ TeV
is plotted in Figure 3, where the shift of the cross section peak
from center of mass energy is clearly seen, especially for large values
of $g_{l}$. This shift is a consequence of ISR and BS.

\begin{figure}
\includegraphics[scale=0.7]{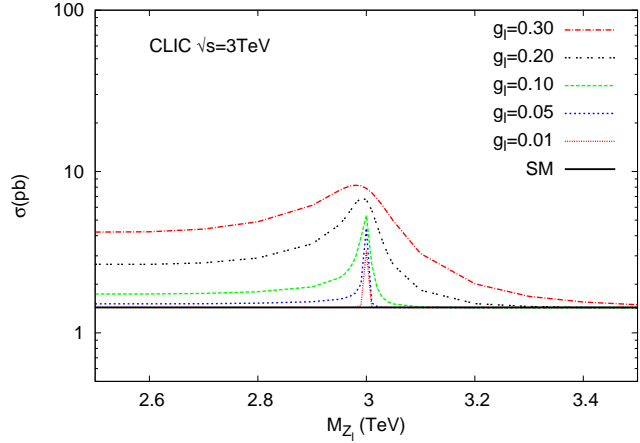}

\caption{Cross section versus $Z_{l}$ mass for different coupling values and
SM background at CLIC with $\sqrt{S}=3$ TeV.}

\end{figure}

In order to show the effects of ISR and BS together with machine design
parameters we present the Figure 4 where cross section versus mass
is plotted for three different cases: $\sqrt{S}=0.5$ TeV without
ISR and BS, ILC with ISR and BS and $\sqrt{S}=0.5$ TeV CLIC with
ISR and BS. It is seen that ISR and BS essentially reduce cross section
at $M_{Z_{l}}\approx\sqrt{S}$, whereas cross section in tailes is
increased by an order.

\begin{figure}
\includegraphics[scale=0.7]{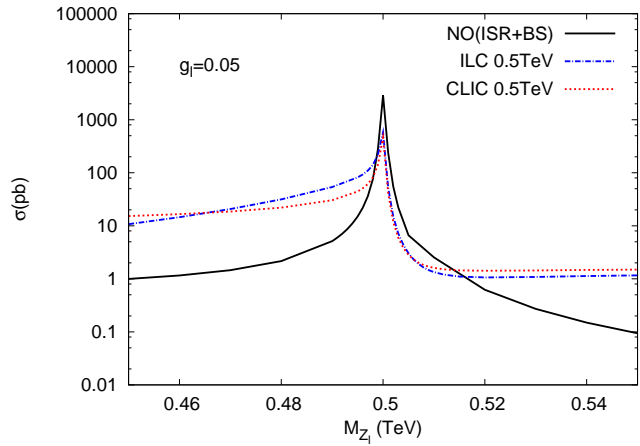}

\caption{ISR and BS effects at ILC and CLIC with $\sqrt{S}=0.5$ TeV.}

\end{figure}

Figure 5 presents effects of ISR and BS depending on coupling constant
$g_{l}$ for ILC and CLIC with $\sqrt{S}=0.5$ TeV. One can see that
these effects essentially reduce corresponding cross section especially
at lower values of $g_{l}$. Furthermore, cross section at ILC exceeds
that of CLIC ($\approx$$25\%$).

\begin{figure}
\includegraphics[scale=0.7]{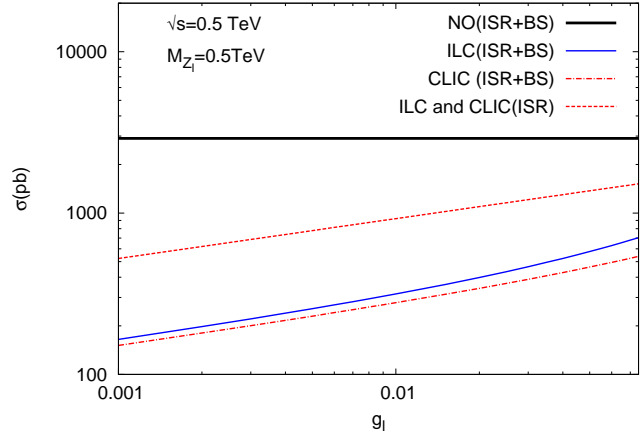}

\caption{ISR and BS effects depending on g$_{l}$ at ILC and CLIC with $\sqrt{S}=0.5$
TeV}

\end{figure}

The ISR and BS effects at CLIC with$\sqrt{S}=3$ TeV are presented
in Figure 6. As expected, ISR and BS effects are more efficient at
higher energies: for $g_{l}=0.05$ reduction factor are 6 and 18 at
$\sqrt{S}=0.5$ TeV and $3$ TeV, respectively.

\begin{figure}
\includegraphics[scale=0.7]{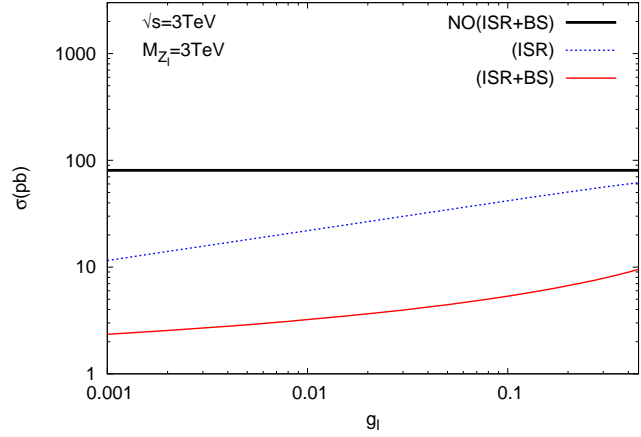}

\caption{ISR and BS effects depending on $g_{l}$ at CLIC with $\sqrt{S}=3$
TeV.}

\end{figure}

In order to determine discovery potential of ILC and CLIC, we have
used following cuts: $\left|M_{inv}(\mu^{+}\mu^{-})-M_{Z_{l}}\right|<10\, GeV$
and $\left|\eta_{\mu}\right|<2$. Statistical significance (S) is
calculated using the following formula:

\begin{equation}
S=\frac{\sigma_{signal}-\sigma_{SM}}{\sqrt{\sigma_{SM}}}\sqrt{L_{int}}\end{equation}

\medskip{}

\begin{figure}
\includegraphics[scale=0.7]{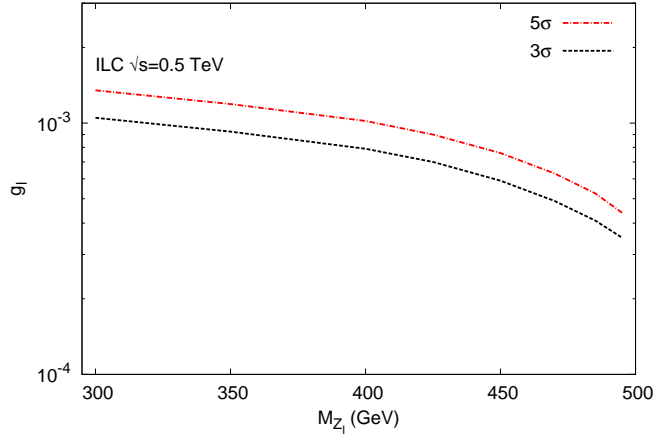}

\caption{Achievable limits for the mass and coupling parameters for $3\sigma$
observations and $5\sigma$ discovery at ILC with $\sqrt{S}=0.5$
TeV.}

\end{figure}

In Figure 7(8) we plot $3\sigma$ and $5\sigma$ contours against
$M_{Z_{l}}$and $g_{l}$ for ILC (CLIC with $\sqrt{S}=0.5$ TeV).
It is seen that both ILC and CLIC will give opportunity to search
leptophilic $Z_{l}$ in the range from $300$ to $500$ GeV down to
$g_{l}\approx10^{-3}$. However, for high mass values the CLIC and
for low mass values the ILC is advantageous. Similar plots for CLIC
with $\sqrt{S}=3$ TeV are given in Figure 9 showing that $Z_{l}$
could be covered up to $M_{Z_{l}}=3$ TeV of $g_{l}\geq10^{-3}$.

\begin{figure}
\includegraphics{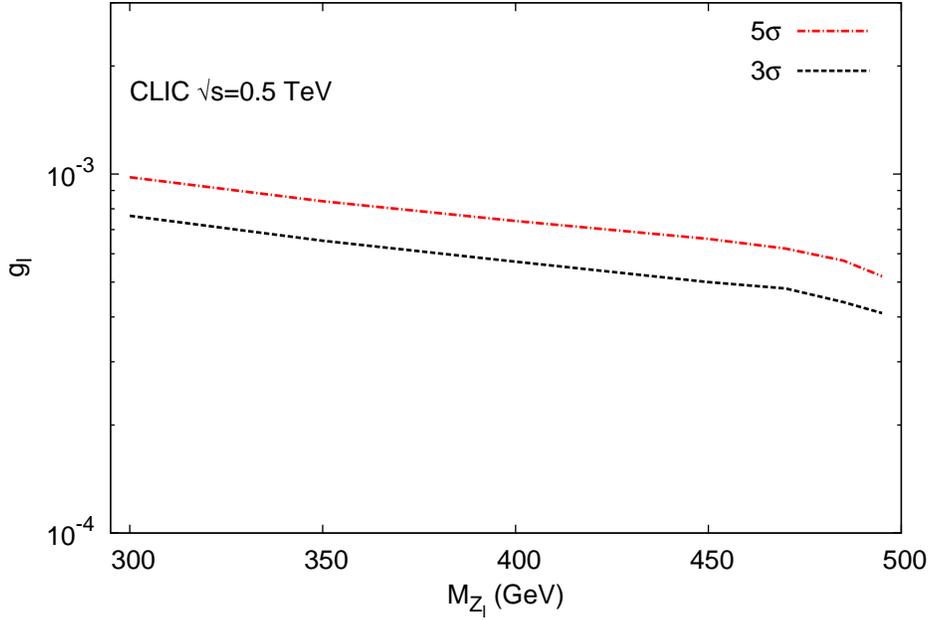}

\caption{Achievable limits for the mass and coupling parameters for $3\sigma$
observations and $5\sigma$ discovery at CLIC with $\sqrt{S}=0.5$
TeV.}

\end{figure}

\begin{figure}
\includegraphics[scale=0.7]{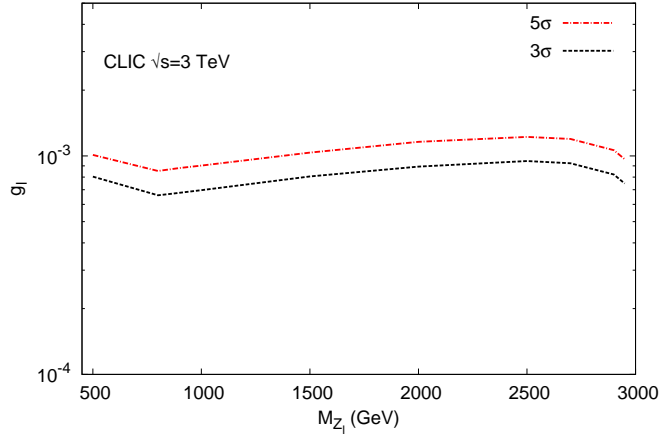}

\caption{Achievable limits for the mass and coupling parameters for $3\sigma$
observations and $5\sigma$ discovery at CLIC with $\sqrt{S}=3$ TeV.}

\end{figure}

In Figure 10 we plotted the invariant mass distribution of final muons
for signal and SM background. It is clear that the signal is well
above the background.

\begin{figure}
\includegraphics[scale=0.7]{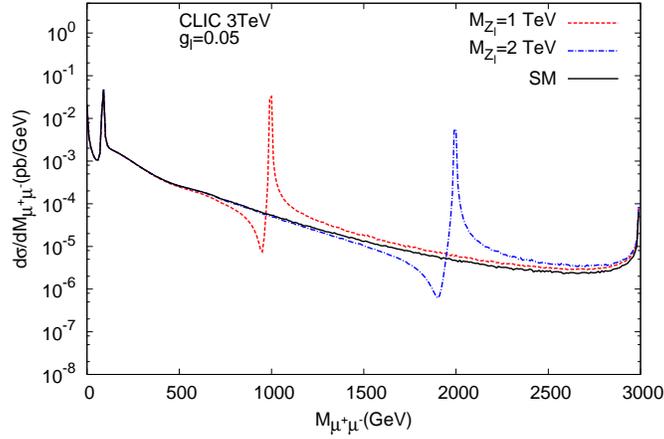}

\caption{Invariant mass distributions of final muons for SM background and
signal. Two different values of M$_{Z_{l}}$ have been used.}

\end{figure}

\section*{4. CONCLUS\.{I}ON}

By investigating the process $e^{+}e^{-}\rightarrow\mu^{+}\mu^{-}$,
we have shown that future linear colliders will give opportunity to
observe leptophilic vector boson with masses up to the center of mass
energy if $g_{l}\geq10^{-3}$. As a result of the calculations done,
we could say that initial state radiation and beamstrahlung will have
important impact for leptophilic $Z_{l}$ vector boson at future linear
colliders. In our calculations we have ignored possible impact of
fourth family leptons on the process when $M_{Z_{l}}>2M_{l_{4}}(M_{\nu_{4}})$.
The work on the subject is ongoing and results will be published elsewhere. 

Finally, massless boson connected to B-L will be considered in separate
paper \cite{Kara}.
\begin{acknowledgments}
This work is supported by DPT and TUBITAK.\end{acknowledgments}

\end{document}